\documentclass[amsmath,amssymb,twocolumn,nofootinbib]{revtex4-1}
\usepackage{graphicx}
\usepackage{epsfig}
\usepackage{color}
\begin{document}
\title{Duality between static spherically or hyperbolically symmetric  solutions and cosmological  solutions in scalar-tensor gravity}
\author{Alexander~Yu.~Kamenshchik}
\email{kamenshchik@bo.infn.it}
\affiliation{Dipartimento di Fisica e Astronomia, Universit\`a di Bologna\\ and INFN,  Via Irnerio 46, 40126 Bologna,
Italy,\\
L.D. Landau Institute for Theoretical Physics of the Russian
Academy of Sciences,\\
 Kosygin str. 2, 119334 Moscow, Russia}
\author{Ekaterina~O.~Pozdeeva}
\email{pozdeeva@www-hep.sinp.msu.ru}
\affiliation{Skobeltsyn Institute of Nuclear Physics, Lomonosov Moscow State University,\\
 Leninskie Gory 1, 119991 Moscow, Russia}
 \author{Alexei A. Starobinsky}
 \email{alstar@landau.ac.ru}
\affiliation{L.D. Landau Institute for Theoretical Physics of the Russian Academy of Sciences,\\
 Kosygin str. 2, 119334 Moscow, Russia,\\
 National Research University Higher School of Economics, 101000 Moscow, Russia}
\author{Alessandro~Tronconi}  
\email{tronconi@bo.infn.it}
\affiliation{Dipartimento di Fisica e Astronomia, Universit\`a di Bologna\\ and INFN, Via Irnerio 46, 40126 Bologna,
Italy}
\author{Tereza Vardanyan}
\email{tereza.vardanyan@bo.infn.it}
\affiliation{Dipartimento di Fisica e Astronomia, Universit\`a di Bologna\\ and INFN, Via Irnerio 46, 40126 Bologna,
Italy}
\author{Giovanni~Venturi}
\email{giovanni.venturi@bo.infn.it}
\affiliation{Dipartimento di Fisica e Astronomia, Universit\`a di Bologna\\ and INFN,  Via Irnerio 46, 40126 Bologna,
Italy}
\author{Sergey~Yu.~Vernov}
\email{svernov@theory.sinp.msu.ru}
\affiliation{Skobeltsyn Institute of Nuclear Physics, Lomonosov Moscow State University,\\
 Leninskie Gory 1, 119991, Moscow, Russia}

\begin{abstract}
We study static spherically  and hyperbolically symmetric solutions of the Einstein equations  in the presence of  a conformally coupled scalar field and compare them
with those in the space filled with a minimally coupled scalar field. We then  study the Kantowski-Sachs cosmological solutions, which are connected with the static solutions by the  duality relations. The main ingredient of these relations is an exchange of roles between the radial and the temporal coordinates, combined with the exchange between the spherical and hyperbolical two-dimensional geometries.  A brief discussion of  questions such as the relation between the Jordan and the Einstein frames and the description of the singularity crossing is also presented.
  \end{abstract}
\maketitle

\section{Introduction}

The exploration of the exact solutions of Einstein equations has been attracting attention of researchers from the dawn of  General Relativity. Exact solutions possessing  spherical symmetry were one of the main branches of this activity
since the time of the classical works by Schwarzschild \cite{Schwarz}, Tolman \cite{Tolman}, Oppenheimer and Volkoff
\cite{O-V}. The study of static spherically symmetric solutions of the Einstein equations in the presence of a massless scalar field has rather a long history \cite{massless1,massless2,massless3,massless4,massless5,massless6,massless7,massless8,Bron2,massless-we,Bronnikov:2006qj,Bronnikov:2016wqp,Galtsov} (see also~\cite{Bronnikov:2018vbs} as a review).
In particular, in paper~\cite{massless-we},     a duality between  spherically symmetric
static solutions in the presence of a massless scalar field and the Kantowski-Sachs cosmological models~\cite{Kant-Sachs},
which instead possess  hyperbolic symmetry was studied. It was noticed also that the spherically symmetric Kantowski-Sachs universes
are connected by a duality transformation to the static solutions possessing  hyperbolic symmetry. In the limiting case
of the absence of the scalar field, the corresponding static solution represents some hyperbolic analogue of the Schwarzschild geometry. While such a hyperbolic solution was mentioned already in paper \cite{Harrison}, its properties were studied in detail in papers \cite{massless-we,massless-we1}. Let us emphasize that the main ingredient of this duality is the exchange of roles between the radial coordinate and the temporal coordinate combined with the exchange between the spherical two-dimensional  geometry and the hyperbolical two-dimensional geometry.  

The study of  gravity models, where a scalar field is non-minimally coupled to the scalar curvature has a long history, too \cite{Jordan,Brans,Wagoner}.
Recently, the actuality of such models has grown due to the study of  inflation models based on the Higgs scalar field non-minimally coupled to gravity
\cite{Higgs-non-min}. Models wherein a non-minimal coupling between gravity and a scalar field is conformal  were also largely studied \cite{Bekenstein, Grib,Frolov,Melnikov1,Melnikov,Linde1979,Starobinsky1981,KPTVV2013,KPTVV2015,KPTVV-sing,we-Bianchi,we-Bianchi1}. These are interesting for two reasons:
firstly, it is relatively easy to find  exact solutions and to establish the relations between these solutions and those obtained in  models with a minimally coupled scalar field; secondly, there is a possibility of a change of sign of the effective gravitational  constant and of the construction of a singularity-free isotropic cosmological model including a scalar field conformally coupled to the scalar curvature. We would also like  to mention the papers~\cite{BarceloVisser,Bron1,Bron2,Bron3,Bron4,Bronnikov:2006qj} where the relation between the exact static solutions in the Jordan frame and in the Einstein frame was studied in detail. In particular, it was noticed that a singularity in one frame can correspond to the regular geometry in another frame. In the paper \cite{Dabrowski} a very detailed compendium of the conformal transformations of such geometrical quantities as the 
metric, Christoffel connection coefficients, the Riemann tensor and different curvature invariants was presented. Further, the conformal transformations of the matter energy-momentum tensor were also given. The connection between the conformal transformations and the duality transformations in superstring theories was explained in the paper \cite{Dabrowski} also. Here, we wish to stress that what we call ``duality'' between static and cosmological solutions in General Relativity is quite different from the duality 
in the superstring theories, because our duality involves the exchanges of the coordinates and not the field variables.   

In the present paper we study static spherically and hyperbolically symmetric geometries in the presence of a massless scalar field conformally coupled to gravity and their relations with Kantowski-Sachs cosmologies. We compare the  solutions found with those obtained in a theory with the minimally coupled massless scalar field.

The structure of the paper is as follows: in the second section we present some general formulas for  gravity with a conformally coupled scalar field. The third section is devoted to the study of static spherically symmetric solutions while in the forth section
we obtain static hyperbolically symmetric solutions. In the fifth section we discuss the duality relations between static and cosmological solutions and present some details concerning time evolution of the Kantowski-Sachs universe in this model. The last section includes some concluding remarks about the relations between different frames and about the problem of the singularity crossing.

\section{Some general formulas for  gravity with a conformally coupled massless scalar field}

Let us consider an action
\begin{equation}
S = \int d^4x \sqrt{-g}\left(U(\sigma)R - \frac12g^{\mu\nu}\sigma_{,\mu}\sigma_{,\nu}\right) .
\label{action}
\end{equation}

The Einstein equations are
\begin{eqnarray}
&&U\left(R_{\mu\nu}-\frac12g_{\mu\nu}R\right)+g_{\mu\nu}\Box U-\nabla_{\mu}\nabla_{\nu} U \nonumber \\
&&=\frac12\sigma_{,\mu}\sigma_{,\nu}-\frac14g_{\mu\nu}\sigma_{,\alpha}\sigma^{,\alpha}.
\label{Einstein}
\end{eqnarray}

The variation with respect to $\sigma$ gives the Klein-Gordon equation
\begin{equation}
\Box \sigma  + \frac{dU}{d\sigma} R =0.
\label{KG}
\end{equation}

The Einstein equations (\ref{Einstein}) can be rewritten as
\begin{eqnarray}
&&U\left(R_{\mu\nu}-\frac12g_{\mu\nu}R\right)+g_{\mu\nu}\frac{d^2U}{d\sigma^2}\sigma_{,\alpha}\sigma^{,\alpha}\nonumber\\
&&{}+g_{\mu\nu}\frac{dU}{d\sigma}\Box \sigma-\frac{d^2U}{d\sigma^2}\sigma_{,\mu}\sigma_{,\nu}-\frac{dU}{d\sigma}\nabla_{\nu}\nabla_{\mu}\sigma \nonumber \\
&&=\frac12\sigma_{,\mu}\sigma_{,\nu}-\frac14g_{\mu\nu}\sigma_{,\alpha}\sigma^{,\alpha}.
\label{Einstein1}
\end{eqnarray}
On contracting Eq.~(\ref{Einstein1}) with the contravariant metric, we get
\begin{eqnarray}
&&{}-UR+3\frac{d^2U}{d\sigma^2}\sigma_{,\mu}\sigma^{,\mu}+3\frac{dU}{d\sigma}\Box\sigma
+\frac12\sigma_{,\mu}\sigma^{,\mu}=0.
\label{Einstein2}
\end{eqnarray}
For the case of a conformal coupling
\begin{equation}
U_c=U_0-\frac{\sigma^2}{12}
\label{conformal}
\end{equation}
one easily finds that
\begin{equation}
R =0
\label{Einstein4}
\end{equation}
and
\begin{equation}
\Box \sigma =0.
\label{KG2}
\end{equation}

\section{Static spherically symmetric solutions}

We shall consider a static spherically symmetric metric in the form
\begin{equation}
ds^2=b^2(r)dt^2-a^2(r)\left(dr^2+d\theta^2+\sin^2\theta d\phi^2\right).
\label{metric}
\end{equation}

For the metric (\ref{metric}) and for the scalar field $\sigma$, that depends only on the radial variable $r$ we obtain
  \begin{equation}
  \label{Boxsigma}
   \Box\sigma(r)={}- \frac{1}{a^2}\sigma'' -\frac{ab'+a'b}{a^3b} \sigma',
  \end{equation}
  where primes mean derivatives with respect to $r$.

  The Ricci scalar  is
\begin{equation}
R={\frac {2   }{
 a^{4}b   }}\left(b''  a^{2}+ b'a' a   -b{a'}^{2}-b a^{2}+2\,b   aa'' \right).
\label{scalar-curv}
\end{equation}

 For the case of the conformal coupling (\ref{conformal}) it follows from Eqs.~(\ref{Einstein4}) and (\ref{scalar-curv}) that
\begin{equation}
    b''  a^{2}+ b'a' a   -b{a'}^{2}-b a^{2}+2baa''=0.
    \label{Requ}
    \end{equation}
    Equation (\ref{KG2}) with (\ref{Boxsigma}) can be easily integrated, giving
\begin{equation}
\label{dsigma}
  \sigma'=\frac{C}{ab},
\end{equation}
where $C$ is an integration  constant.

The Einstein equations are now
\begin{equation}\label{Eq00mm}
6U_c\left[{a'}^2+a^2-2aa''\right]=\frac{C^2}{2b^2}+
\frac{C\sigma a b'}{b^2}\,,
\end{equation}
\begin{equation}\label{Eq11mm}
6U_c\left[a^2b-{a'}^2b-2a'ab'\right]={}-
Ca\sigma\left[\frac{b'}{b}+2\frac{a'}{a}\right]-\frac{3C^2}{2b}\,,
\end{equation}
\begin{equation}\label{Eq22mm}
6U_c\left[{a'}^2b-aba''-a^2b''\right]=\frac{C^2}{2b}+\sigma C a'\,.
\end{equation}

In order to simplify the equations obtained above we introduce new functions
\begin{equation}\label{aebe}
    a_e(r)=\sqrt{1-\frac{\sigma(r)^2}{12U_0}}a(r),\quad b_e(r)=\sqrt{1-\frac{\sigma(r)^2}{12U_0}}b(r).
\end{equation}
In terms of these functions, Eqs.~(\ref{Eq00mm})--(\ref{Eq22mm}) take the following form
\begin{equation}
\label{equ00e}
a_e^2b_e^2-2b_e^2a_e a_e''+b_e^2{a_e'}^2=\frac{C^2}{4U_0},
\end{equation}
\begin{equation}
\label{equ11e}
  a_e^2b_e^2-b_e^2{a_e'}^2-2b_ea_ea_e'b_e'=-\frac{C^2}{4U_0},
\end{equation}
\begin{equation}
\label{equ22e}
b_e^2a_e a_e''-b_e^2{a_e'}^2+b_e a_e^2b_e''=-\frac{C^2}{4U_0}
\end{equation}

Equation~(\ref{Requ}) in terms of the new variables is
\begin{equation}
\label{equRe}
  2b_e^2a_ea_e'' -a_e^2b_e^2-b_e^2 {a_e'}^2+b_ea_e^2b_e''+b_ea_e a_e' b_e'=-\frac{C^2}{4U_0}.
\end{equation}

The introduction  of the new functions $a_e$ and $b_e$ used together  with Eq.~(\ref{dsigma}) allows us to obtain equations independent of the scalar field  $\sigma$ and its derivatives.

On introducing
\begin{equation*}
  A=\frac{a_e'}{a_e},\qquad B\equiv\frac{b_e'}{b_e},
\end{equation*}
we  can rewrite Eqs.~(\ref{equ00e})--(\ref{equ22e})  in the following form:
\begin{equation}
\label{equ00ee}
    1-2A'-A^2=\frac{C^2}{4U_0b_e^2a_e^2},
    \end{equation}
    \begin{equation}
\label{equ11ee}
    1-2AB-A^2={}-\frac{C^2}{4U_0b_e^2a_e^2},
    \end{equation}
    \begin{equation}
\label{equ22ee}
    A'+B'+B^2={}-\frac{C^2}{4U_0b_e^2a_e^2}.
    \end{equation}
The resulting equations are quite similar  to the Einstein equations for the model with minimally coupled massless scalar field, considered in~\cite{massless-we}.

On summing Eqs.~(\ref{equ00ee}) and (\ref{equ11ee}) and Eqs.~(\ref{equ00ee}) and (\ref{equ22ee}), we obtain the following equations:
\begin{equation}
1-A-A'-AB-A^2=0,
\label{eq1}
\end{equation}
\begin{equation}
1-A'+B'+B^2-A^2=0.
\label{eq2}
\end{equation}
From this pair of equations one can obtain  another:
\begin{equation}
A=-\frac{B'}{B}-B,
\label{eq3}
\end{equation}
\begin{equation}
\left(\frac{1}{B}\right)''-\frac{1}{B} = 0.
\label{eq4}
\end{equation}
There are two independent solutions of Eq.~(\ref{eq4}). One of these solutions is proportional to the hyperbolic cosine and the other is proportional to the hyperbolic sine.
Let us choose  as a solution
\begin{equation}
B = \frac{\gamma}{\cosh r},
\label{B}
\end{equation}
where $\gamma$ is a constant.
Then from Eq.~(\ref{eq3}) it follows that
\begin{equation}
A=\tanh r - \frac{\gamma}{\cosh r}.
\label{A}
\end{equation}
On substituting  the expression (\ref{A}) into the left-hand side of Eq.~(\ref{equ00ee}), we see that it is equal to $-\frac{(1+\gamma^2)}{\cosh^2r} < 0$, while the right-hand side of this equation is positive. Thus, we should discard the solution (\ref{B})--(\ref{A}).

Let us now consider
\begin{equation}
B = \frac{\gamma}{\sinh r},
\label{B1}
\end{equation}
 then
 \begin{equation}
 A = \coth r -\frac{\gamma}{\sinh r}.
 \label{A1}
 \end{equation}
 On substituting the expression (\ref{A1}) into Eq.~(\ref{equ00ee}), we obtain
 \begin{equation}
 \frac{1-\gamma^2}{\sinh^2r}=\frac{C^2}{4U_0b_e^2a_e^2},
 \label{C}
 \end{equation}
which
tells us that
\begin{equation}
\gamma^2 \leqslant 1.
\label{gamma}
\end{equation}

On now integrating Eqs.~(\ref{B1}) and (\ref{A1}), we obtain
\begin{equation}
b_e=b_0\left(\tanh \frac{r}{2}\right)^{\gamma}
\label{be}
\end{equation}
and
\begin{equation}
a_e = a_0\frac{\sinh r}{\left(\tanh \frac{r}{2}\right)^{\gamma}},
\label{ae}
\end{equation}
where $a_0$ and $b_0$ are constants.

On substituting the expressions (\ref{be}) and (\ref{ae}) into Eq.~(\ref{C}), we obtain
\begin{equation}
C=\pm 2a_0b_0\sqrt{U_0}\sqrt{1-\gamma^2}.
\label{C1}
\end{equation}
In what follows we shall choose the ``plus" sign  on the right-hand side of Eq.~(\ref{C1}) without loss of  generality.

On substituting (\ref{C1}) together with (\ref{ae}) and (\ref{be}) and (\ref{aebe}) into Eq.~(\ref{dsigma}), we obtain
\begin{equation}
\frac{\sigma'}{1-\frac{\sigma^2}{12U_0}} = \frac{2\sqrt{U_0}\sqrt{1-\gamma^2}}{\sinh r}.
\label{sigma-eq}
\end{equation}
On integrating this equation, we obtain
\begin{equation}
\sigma = \sqrt{12U_0}\frac{A_0\left(\tanh \frac{r}{2}\right)^{2\sqrt{\frac{1-\gamma^2}{3}}}-1}{A_0\left(\tanh \frac{r}{2}\right)^{2\sqrt{\frac{1-\gamma^2}{3}}}+1},
\label{sigma-sol}
\end{equation}
where $A_0 > 0$ is an integration constant.
On using the solution (\ref{sigma-sol}) we can finally write down  the expressions for the functions $a$ and $b$:
\begin{equation}
a=\frac{a_0\left(A_0\left(\tanh \frac{r}{2}\right)^{2\sqrt{\frac{1-\gamma^2}{3}}}+1\right)\sinh r}
{2\sqrt{A_0}\left(\tanh \frac{r}{2}\right)^{\gamma + \sqrt{\frac{1-\gamma^2}{3}}}},
\label{a-final}
\end{equation}
\begin{equation}
b= \frac{b_0\left(A_0\left(\tanh \frac{r}{2}\right)^{2\sqrt{\frac{1-\gamma^2}{3}}}+1\right)\left(\tanh \frac{r}{2}\right)^{\gamma}}
{2\sqrt{A_0}\left(\tanh \frac{r}{2}\right)^{\sqrt{\frac{1-\gamma^2}{3}}}}.
\label{b-final}
\end{equation}

Let us first look at the particular cases, when $\gamma = \pm 1$. For these cases the derivatives of the scalar field are equal to zero (see Eq.~(\ref{C1})) and the case is equivalent to the case of the empty space. The presence of a constant scalar field
$\sigma = \sqrt{12U_0}\frac{A_0-1}{A_0+1}$ implies simply some changes of the Newton constant. Thus, this case coincides with that considered for a minimally coupled scalar field in paper~\cite{massless-we}.  Let us  give here some details for completeness. In the case $\gamma =1$, the metric (\ref{metric}) has the form
\begin{eqnarray}
&&ds^2=\frac{b_0^2(A_0+1)^2\tanh^2\frac{r}{2}}{4A_0}dt^2\nonumber \\
&&{}-\frac{a_0^2(A_0+1)^2\cosh^4\frac{r}{2}}{A_0}(dr^2+d\theta^2+\sin^2\theta d\phi^2).
\label{Sch}
\end{eqnarray}
On introducing a new ``Schwarzschild'' radial variable
\begin{equation}
\tilde{r} = \frac{a_0(A_0+1)}{\sqrt{A_0}}\cosh^2\frac{r}{2},
\label{Sch1}
\end{equation}
we can rewrite the metric (\ref{Sch}) in the familiar Schwarzschild form
\begin{eqnarray}
&&ds^2 = \frac{b_0^2(A_0+1)^2}{4A_0}\left(1-\frac{a_0(A_0+1)}{\sqrt{A_0}}\frac{1}{\tilde{r}}\right)dt^2\nonumber \\
&&{}-\frac{d\tilde{r}^2}{\left(1-\frac{a_0(A_0+1)}{\sqrt{A_0}}\frac{1}{\tilde{r}}\right)}-\tilde{r}^2(d\theta^2+\sin^2\theta d\phi^2),
\label{Sch2}
\end{eqnarray}
where the quantity $\frac{a_0(A_0+1)}{\sqrt{A_0}}$ plays the role of the Schwarzschild radius and
the constant $\frac{b_0^2(A_0+1)^2}{4A_0}$ can be absorbed in the definition of the time parameter.

For the case $\gamma = -1$, the metric (\ref{metric}) has the form
\begin{eqnarray}
&&ds^2=\frac{b_0^2(A_0+1)^2\coth^2\frac{r}{2}}{4A_0}dt^2\nonumber \\
&&{}-\frac{a_0^2(A_0+1)^2\sinh^4\frac{r}{2}}{A_0}(dr^2+d\theta^2+\sin^2\theta d\phi^2).
\label{Sch3}
\end{eqnarray}
On introducing a variable $\hat{r}$ by
\begin{equation}
\hat{r} =\frac{a_0(A_0+1)}{\sqrt{A_0}}\sinh^2\frac{r}{2},
\label{Sch4}
\end{equation}
we can rewrite the metric (\ref{Sch3}) as
\begin{eqnarray}
&&ds^2 = \frac{b_0^2(A_0+1)^2}{4A_0}\left(1+\frac{a_0(A_0+1)}{\sqrt{A_0}}\frac{1}{\hat{r}}\right)dt^2\nonumber \\
&&{}-\frac{d\hat{r}^2}{\left(1+\frac{a_0(A_0+1)}{\sqrt{A_0}}\frac{1}{\hat{r}}\right)}-\tilde{r}^2(d\theta^2+\sin^2\theta d\phi^2),
\label{Sch5}
\end{eqnarray}
where on choosing $a_0 < 0$, we again have the standard Schwarzschild metric, where the Schwarzschild radius is proportional to some positive point-like mass.

On now, looking at the expressions (\ref{a-final}) and (\ref{b-final}), we get
\begin{equation}
b(r) \sim  r^{\gamma-\sqrt{\frac{1-\gamma^2}{3}}},\,
\qquad a(r) \sim  r^{1-\gamma-\sqrt{\frac{1-\gamma^2}{3}}},
\label{ab-as}
\end{equation}
when  $r \rightarrow 0$.

We can see that there is another special value of the parameter
$\gamma$, it is
\begin{equation}
\gamma = \frac12.
\label{special}
\end{equation}
Indeed, if $\gamma = 1/2$, then at $r=0$ both  factors $a(r)$ and $b(r)$ and, hence, the corresponding metric coefficients are finite. This regime does not have a counterpart in the case of a minimally coupled scalar field~\cite{massless-we} and we shall discuss it in detail later.

It is easy to see that for $\gamma > 1/2$,
\begin{equation*}
b(r) \rightarrow 0,
\quad
a(r)  \rightarrow \infty,\qquad\ {\rm when}\ r \rightarrow 0.
\end{equation*}
If $\gamma < 1/2$, then the behavior of the functions $a$ and $b$ is the opposite: $b(r) \rightarrow \infty$, while $a(r) \rightarrow 0$, when
$r \rightarrow 0$. A simple calculation shows that if $\gamma \neq 1/2$, then at $r \rightarrow 0$ the invariant
\begin{equation}
R_{\mu\nu}R^{\mu\nu} \sim B_0r^{-4\left(2-\gamma-\sqrt{\frac{1-\gamma^2}{3}}\right)} \rightarrow \infty,
\label{Ricci}
\end{equation}
where $B_0$ is a positive constant. Thus, the solutions with $\gamma \neq 1/2$ contain a real singularity at $r=0$.
In this case we consider $a(r)$ and $b(r)$ for $r\geqslant 0$ only.

Such a singularity is absent for the case when $\gamma = 1/2$, because the functions $a$ and $b$ are finite at $r=0$. The explicit expression for the metric is now
\begin{widetext}
\begin{eqnarray}
&&ds^2=\frac{b_0^2\left(A_0\tanh\frac{r}{2}+1\right)^2}{4A_0}dt^2
-\frac{a_0^2\left[A_0\tanh\frac{r}{2}+1\right]^2\cosh^4\frac{r}{2}}{A_0}(dr^2+d\theta^2+\sin^2\theta d\phi^2).
\label{special1}
\end{eqnarray}
\end{widetext}
The scalar field is given by
 \begin{equation}
 \sigma = \sqrt{12U_0}\frac{A_0\tanh\frac{r}{2}-1}{A_0\tanh\frac{r}{2}+1}.
 \label{special2}
 \end{equation}
 One can see that both the expressions (\ref{special1}) and (\ref{special2}) are quite regular at $r=0$ and can be smoothly continued in the
 region $r < 0$. The expression for the scalar field in this region is such that $\sigma^2 > 12U_0$ and, hence, $U_c < 0$ and we enter into the antigravity regime, without crossing any singularity. Let us note  that the effect of the disappearance of the singularity due to the non-minimal coupling is connected with a high symmetry of the geometry (here, it being the spherical symmetry). A similar effect 
was observed also in paper \cite{KPTVV-sing} for a flat Friedmann universe. On the other hand, as was shown in paper 
\cite{Starobinsky1981} and then studied in some details in paper \cite{we-Bianchi1} for the case of the Bianchi-I universe the transition to the regime of antigravity, where $U_c < 0$, is accompanied by the appearance of the cosmological singularity.
Thus, one should expect that removing the assumption of
spherical symmetry or, more generally, axial symmetry, will result in
formation of general curvature singularities just on spatial hypersurfaces
beyond which gravity becomes repulsive. 

 Let us see what happens at $r < 0$. There are two options. If the integration constant $A_0 < 1$, then the geometry is regular for all  values of the variable $r$. The asymptotic expression for the metric (\ref{special1}) at $r \rightarrow -\infty$ is
 \begin{eqnarray}
 &&ds^2=\frac{b_0^2(1-A_0)^2}{4A_0}dt^2\nonumber \\
 &&-\frac{a_0^2(1-A_0)^2}{16A_0}e^{-2r}(dr^2+d\theta^1+\sin^2\theta d\phi^2).
 \label{special3}
 \end{eqnarray}
 On introducing a variable
\begin{equation}
\bar{r}=\frac{a_0(1-A_0)e^{-r}}{4\sqrt{A_0}},
\label{special4}
\end{equation}
we can rewrite the metric (\ref{special3}) as
\begin{equation}
ds^2=\frac{b_0^2(1-A_0)^2}{4A_0}dt^2-d\bar{r}^2-\bar{r}^2(d\theta^2+\sin^2\theta d\phi^2),
\label{special5}
\end{equation}
and it describes the Minkowski spacetime. It is easy to see that at $r \rightarrow \infty$ we again encounter an asymptotically flat Minkowski spacetime.
Let us note that for a value of the radial variable
\begin{equation}
r ={} -{\rm arctanh} A_0,
\label{throat}
\end{equation}
the  factor $a(r)$ has a minimum value. Thus, we can imagine that this value of the variable $r$ corresponds to a throat of some wormhole-like configuration.\footnote{Note that the fact that this configuration requires the ghost
behavior of graviton in the antigravity regime for $r < 0$ is in the
agreement with the general theorem proved in Ref. \cite{Bron-Star} that there
are no non-singular wormholes in scalar-tensor gravity without
ghosts.}

Let us again  emphasize   that the absence  of the singularity at $r=0$ provided $\gamma = 1/2$ is connected with the presence of the conformal coupling in our model. Indeed, on making the transition from the Jordan frame to the Einstein frame, where the coupling becomes minimal, one may encounter the singularity at
$r=0$. Therefore, in this case the conformal continuation, described in~\cite{Bronnikov:2006qj}, is possible. A similar phenomenon for the Friedmann-Lema\^itre-Robertson-Walker cosmology was described in detail in paper~\cite{KPTVV-sing}.

Let us now consider a more interesting case wherein the integration constant $A_0 \geq 1$. When
\begin{equation}
r \rightarrow r_0 = -2{\rm arctanh}\frac{1}{A_0},
\label{special6}
\end{equation}
both  scale factors tend to zero as $(r-r_0)$ and we stumble upon the singularity, characterized by the invariant
\begin{equation}
R_{\mu\nu}R^{\mu\nu} \sim \frac{1}{(r-r_0)^8}.
\label{special7}
\end{equation}
 Nevertheless, for $r < r_0$ the metric and the scalar field are well defined and one can construct  the continuation of the solution into this region.
 Then, for $r \rightarrow -\infty$ we again have an asymptotically flat Minkowski spacetime.
We wish to note that in contrast with the case of  $r=0$ the singularity at $r=r_0$ and $A_0 \geq 1$ arises due to the presence of the conformal coupling. Indeed, the transformation to the Einstein frame eliminates this singularity.

Let  end  this section by observing  that for the case $\gamma \neq 1/2$ the continuation of the solutions
``beyond the singularity'' looks rather problematic, even if formally the corresponding equations are satisfied. The point is that the  function  $\tanh \frac{r}{2}$ enters into the solutions in  powers  of  $\gamma$ and $\sqrt{\frac{1-\gamma^2}{3}}$  and is ill-defined
at $r<0$ when $\tanh \frac{r}{2}$ is negative.

\section{Static hyperbolically symmetric solutions for the case of a conformally coupled scalar field}

We shall consider a static hyperbolically symmetric metric of the form \cite{massless-we,massless-we1}.
\begin{equation}
ds^2=b^2(r)dt^2-a^2(r)\left(dr^2+d\chi^2+\sinh^2\chi d\phi^2\right),
\label{metric-hyp}
\end{equation}
where the hyperbolic angle $\chi$  runs from $0$ to $\infty$.
All  considerations are analogous to those presented in the preceding section.

We then obtain the general solution in the following form;
\begin{widetext}
\begin{eqnarray}
&&ds^2 = \frac{b_0^2\left(A_0\left(\tan\frac{r}{2}\right)^{2\sqrt{\frac{1-\gamma^2}{3}}}+1\right)^2\left(\tan\frac{r}{2}\right)^{2\gamma}}
{4A_0\left(\tan\frac{r}{2}\right)^{2\sqrt{\frac{1-\gamma^2}{3}}}}dt^2
-\frac{a_0^2\left(A_0\left(\tan\frac{r}{2}\right)^{2\sqrt{\frac{1-\gamma^2}{3}}}+1\right)^2\sin^2 r}
{4A_0\left(\tan\frac{r}{2}\right)^{2\sqrt{\frac{1-\gamma^2}{3}}+2\gamma}}(dr^2+d\chi^2+\sinh^2\chi d\phi^2),\,\,\,
\label{hyp}
\end{eqnarray}
\end{widetext}
while
\begin{equation}
\sigma = \sqrt{12U_0}\frac{A_0\left(\tan\frac{r}{2}\right)^{2\sqrt{\frac{1-\gamma^2}{3}}}-1}
{A_0\left(\tan\frac{r}{2}\right)^{2\sqrt{\frac{1-\gamma^2}{3}}}+1}.
\label{hyp1}
\end{equation}

For the cases $\gamma = \pm 1$ we obtain the pseudo-Schwarzschild solution, which was mentioned in the paper by Harrison \cite{Harrison} as
``degenerate solution III-9'' and whose properties were studied in detail in papers \cite{massless-we,massless-we1}.
If $\gamma \neq 1/2$,  at the point $r=0$ one has the singularity of the same kind as that studied in the preceding section.
However, another singularity arises at $r = \pi$, if $\gamma \neq -1/2$. In this case the invariant $R_{\mu\nu}R^{\mu\nu}$ behaves as
\begin{equation}
R_{\mu\nu}R^{\mu\nu} \sim (\pi -r)^{-4\left(2+\gamma -\sqrt{\frac{1-\gamma^2}{3}}\right)}.
\label{new-sing}
\end{equation}
Note that the right-hand sides of Eqs. (58) and (62) have the
standard $\rho^{-4}$  behavior if expressed in terms of the proper
distance $\rho$.
Thus, if $\gamma \neq \pm 1/2$ then, the solution (\ref{hyp}), (\ref{hyp1}) is well defined between two singularities at $r=0$ and $r=\pi$.
Let us now consider  two particular cases. If $\gamma = 1/2$,  the solution (\ref{hyp}), (\ref{hyp1}) has the following form:
\begin{widetext}
\begin{eqnarray}
&&ds^2 = \frac{b_0^2 \left(A_0\tan\frac{r}{2}+1\right)^2}{4A_0}dt^2-\frac{a_0^2\left[4A_0\tan\frac{r}{2}+1\right]^2\sin^2r}{A_0\tan^2\frac{r}{2}}\left[dr^2+d\chi^2+\sinh^2\chi d\phi^2\right].
\label{hyp2}
\end{eqnarray}
\end{widetext}
\begin{equation}
\sigma = \sqrt{12U_0}\frac{A_0\tan\frac{r}{2}-1}{A_0\tan\frac{r}{2}+1}.
\label{hyp3}
\end{equation}
The value $r=0$ is now  regular and we can construct a continuation of the solution into the region $r < 0$. However, in  this region we encounter
a singularity at
\begin{equation}
r_1 = -2{\rm arctan}\frac{1}{A_0}.
\label{hyp4}
\end{equation}
One can construct a continuation through this singularity because the expressions (\ref{hyp2}) and (\ref{hyp3}) are well defined at $r < r_1$. Finally, we encounter the singularity, which was already described above at $r = -\pi$. Thus, one can say that the solutions (\ref{hyp2}), (\ref{hyp3}) are defined between the two singularities at $r = -\pi$ and $r = +\pi$ with an intermediate singularity at $r=r_1=-2{\rm arctan}\frac{1}{A_0}$,  which can be continued through.

Let us consider another particular case $\gamma = -1/2$. The solution is now
\begin{widetext}
\begin{eqnarray}
&&ds^2=\frac{b_0^2\left(A_0\tan\frac{r}{2}+1\right)^2}{4A_0\tan^2\frac{r}{2}}dt^2-\frac{a_0^2\left(A_0\tan\frac{r}{2}+1\right)^2\sin^2r}{4A_0}
(dr^2+d\chi^2+\sinh^2\chi d\phi^2).
\label{hyp5}
\end{eqnarray}
\end{widetext}
The expression for the scalar $\sigma$ is given by Eq.~(\ref{hyp3}).
The solution (\ref{hyp5}) is defined between two singularities at $r = 0$ and $r=2\pi$ and is nonsingular at $r=\pi$. There is also
an intermediate singularity at $r = 2\pi - 2{\rm arctan}\frac{1}{A_0}$.

\section{Relation between static and cosmological solutions}
In this section we shall use the method for the construction of  cosmological solutions, starting from the duality relations described in the paper \cite{massless-we}.
The corresponding  transformations can be considered as a special kind of complex
transformations used for the construction of new solutions of the Einstein equations (see e.g. \cite{exact}). As an example one can mention also the complex transformations 
connecting cosmological Kasner solutions \cite{Kasner} for a Bianchi-I universe with the static Kasner solutions (see, e.g. \cite{Harvey}).  However, the particular form of the complex transformations,
exchanging hyperbolic and spherical symmetry in the form implemented in paper \cite{massless-we} and in the present paper does not appear to be widely
used.
Let us consider the static spherically symmetric spacetime. If we make the substitution
\begin{equation}
r \leftrightarrow t,
\label{dual}
\end{equation}
followed by a change of the sign of all the metric components
\begin{equation}
g_{\mu\nu} \rightarrow -g_{\mu\nu},
\label{dual1}
\end{equation}
and by the substitution
\begin{equation}
\theta \rightarrow i\chi,
\label{dual2}
\end{equation}
we obtain a Kantowski-Sachs cosmological solution, where the spherical symmetry is replaced by the hyperbolic one:
\begin{widetext}
\begin{equation}
ds^2=\frac{a_0^2\left(A_0\left(\tanh\frac{t}{2}\right)^{2\sqrt{\frac{1-\gamma^2}{3}}}+1\right)^2\sinh^2t}{4A_0\left(\tanh\frac{t}{2}\right)^{2\gamma+2\sqrt{\frac{1-\gamma^2}{3}}}}
(dt^2-d\chi^2-\sinh^2\chi d\phi^2)-\frac{b_0^2\left(A_0\left(\tanh\frac{t}{2}\right)^{2\sqrt{\frac{1-\gamma^2}{3}}}+1\right)^2\left(\tanh\frac{t}{2}\right)^{2\gamma-2\sqrt{\frac{1-\gamma^2}{3}}}}{4A_0}dr^2.
\label{KS}
\end{equation}
It is curious to look at the particular solution for the case $\gamma = 1/2$ and $A_0 < 1$. Now
\begin{equation}
ds^2=\frac{a_0^2\left(A_0\tanh\frac{t}{2}+1\right)^2\cosh^4 \frac{t}{2}}{A_0}(dt^2-d\chi^2-\sinh^2\chi d\phi^2)-\frac{b_0^2\left(A_0\tanh\frac{t}{2}+1\right)^2}{4A_0}dr^2.
\label{KS1}
\end{equation}
\end{widetext}
One can see that the evolution of the universe is a non-singular one. The scale corresponding to the variable $r$ (which can be both compact or non-compact) is almost
constant. Let us look at the evolution of the two-dimensional hyperboloid with the metric
$$
d\chi^2 + \sinh^2\chi d\phi^2.
$$
When $t \rightarrow -\infty$ the metric of the universe can be represented as
\begin{equation}
ds^2=d\tilde{t}^2-\tilde{t}^2(d\chi^2 + \sinh^2\chi d\phi^2)-\frac{b_0^2(1-A_0)^2}{4A_0}dr^2,
\label{KS2}
\end{equation}
where a cosmic time parameter $\tilde{t}$ is defined by
\begin{equation}
\tilde{t} = -\frac{a_0(1-A_0)e^{-t}}{\sqrt{A_0}}.
\label{KS3a}
\end{equation}
It is easy to see that the metric (\ref{KS2}) describes the direct product of the line or circle by the $2+1$ dimensional Milne universe, which is equivalent to the Minkowski spacetime. An analogous expression can be written for $t \rightarrow +\infty$. Thus, the universe begins its evolution in the distant past from the asymptotically Minkowski spacetime,
represented in the Milne form, then it contracts until the moment $t_2=-2{\rm arctanh} A_0$ and the it begins an expansion, which ends again in the asymptotically flat Minkowski spacetime.

Another interesting case arises if  we start from the static hyperbolically symmetric metric (\ref{hyp}) and make the duality transformations presented above with the difference that now
\begin{equation*}
\chi \rightarrow i\theta.
\end{equation*}

Then we arrive at the Kantowski-Sachs universe, where the spatial sections are direct products of the one-dimensional submanifold (the $r$ variable) and a two-dimensional sphere:
\begin{widetext}
\begin{equation}
ds^2 = \frac{a_0^2\left(A_0\left(\tan\frac{t}{2}\right)^{2\sqrt{\frac{1-\gamma^2}{3}}}+1\right)^2\sin^2 t}{4A_0\left(\tan\frac{t}{2}\right)^{2\sqrt{\frac{1-\gamma^2}{3}}+2\gamma}}
(dt^2-d\theta^2-\sin^2\theta d\phi^2)-\frac{b_0^2\left(A_0\left(\tan\frac{t}{2}\right)^{2\sqrt{\frac{1-\gamma^2}{3}}}+1\right)^2\left(\tan\frac{t}{2}\right)^{2\gamma-2\sqrt{\frac{1-\gamma^2}{3}}}}{4A_0}dr^2.
\label{KS3}
\end{equation}
\end{widetext}

As before the cases $\gamma = \pm 1$ describe an empty Kantowski-Sachs universe and they are well known.

If $\frac12< \gamma< 1$, then one has the singularities at $t=0$ and $t=\pi$ and it is not clear if the continuation through these singularities makes sense. When $t \rightarrow 0$, $a(t) \rightarrow \infty$, while $b(t) \rightarrow 0$. At $t \rightarrow \pi$,
the scale factor $a$ tends to zero, while $b \rightarrow \infty$. All the evolution takes place at the gravity regime ($U_c \geq 0$).

If $-\frac12 < \gamma < 1/2$, then at $t = 0$ we have a singularity such that $a \rightarrow 0$ and
$b \rightarrow \infty$. Then at $t \rightarrow \pi$ the scale factor $a$ again vanishes while $b$ grows infinitely.

If $-1 < \gamma < -1/2$, then at $t \rightarrow 0$, $a \rightarrow 0$ while $b \rightarrow \infty$. When
$t \rightarrow \pi$, $a \rightarrow \infty$ and $b \rightarrow  0$.

Let us consider in detail the particular cases $\gamma = \pm 1/2$.
For $\gamma = 1/2$ the metric is given by
\begin{widetext}
\begin{equation}
ds^2=\frac{a_0^2\left(A_0\tan\frac{t}{2}+1\right)^2\cos^4 \frac{t}{2}}{A_0}
(dt^2-d\theta^2-\sin^2\theta d\phi^2)-\frac{b_0^2\left(A_0\tan\frac{t}{2}+1\right)^2}{4A_0}dr^2.
\label{KS4}
\end{equation}
\end{widetext}
This metric is regular at $t=0$ and has  singularities at $t=\pm \pi$ and at $t = t_0 = -2{\rm arctan}\frac{1}{A_0}$.
At $t \rightarrow  \pm \pi$ the scale factor $a \rightarrow 0$ while $b \rightarrow \infty$. At $t \rightarrow t_0$ both  scale factors vanish.
At $t < 0$, we find ourselves in the region with antigravity because $U_c < 0$. We see that the expression (\ref{KS4}) contains
only integer powers of the trigonometrical functions and one can describe  the crossing of the singularities in a unique way.
Thus, we can imagine an infinite periodic evolution of the universe. Let us consider a period between $-\pi$ and $\pi$. At $t = -\pi$,
the universe goes out of the singularity with the vanishing value of the scale factor $a$ and an infinite value of the scale factor $b$.
Then, the scale factor $b$ decreases and vanishes when the universe approaches to the singularity at $t = t_0 = -2{\rm arctan}\frac{1}{A_0}$. Meanwhile the scale factor $a$ increases and reaches its maximal value at $t=t_1 = -\pi +{\rm arctan}A_0$ and then begins decreasing and vanishes at the singularity at $t=t_0$. After that $b$ increases reaching an infinite value at the singularity at $t=\pi$, while $a$ increases until $t = t_2 ={\rm arctan}A_0$, where it achieves its maximum value and then decreases and vanishes
at $t = \pi$. Then, the evolution repeats itself.
Let us now also look more carefully to the structure of the anisotropy of these cosmological singularities. In the vicinity of the moment $t \rightarrow \pi$, the asymptotic expressions for the metric coefficients become simpler and we can introduce a cosmic time parameter $T \rightarrow 0$. The metric now has the following form: 
\begin{equation}
ds^2 = dT^2 -c_1^2Td\theta^2 -c_2^2T\sin^2\theta d\phi^2 - c_3\frac{1}{T}dr^2.
\label{Kasner}
\end{equation}
This form has a structure similar to that of the Kasner solution for a Bianchi-I universe \cite{Kasner,Lif-Khal},
where the Kasner indices have the values
\begin{equation}
p_1=\frac12,\ p_2 = \frac12,\ p_3=-\frac12.
\label{Kasner1}
\end{equation}
Let us note  that while these indices do not satisfy the standard Kasner relations \cite{Kasner,Lif-Khal}
\begin{equation}
p_1+p_2+p_3=p_1^2+p_2^2+p_3^2=1,
\label{Kasner2}
\end{equation}
they satisfy the generalized relation
\begin{equation}
\sum _{i=1}^3p_i^2=2\sum_{i=1}^3p_i-\left(\sum_{i=1}^3p_i\right)^2,
\label{Kasner3}
\end{equation}
discussed in our preceding paper \cite{we-Bianchi1}. We can find a similar asymptotic representation of the metric 
(\ref{KS4}) in the vicinity of the singularity at $t=t_0$. It is 
\begin{equation}
ds^2 = dT^2 -c_1^2Td\theta^2 -c_2^2T\sin^2\theta d\phi^2 - c_3T dr^2.
\label{Kasner4}
\end{equation}
 Thus, this behavior is isotropic and the Kasner indices 
 \begin{equation}
 p_1=p_2=p_3=\frac12
 \label{Kasner5}
 \end{equation}
 again satisfy the relation (\ref{Kasner3}).

Lets us consider another particular case where $\gamma = -1/2$. The metric is now
\begin{widetext}
\begin{equation}
ds^2=\frac{a_0^2\left(A_0\tan\frac{t}{2}+1\right)^2\sin^2t}{4A_0}(dt^2-d\theta^2-\sin^2\theta d\phi^2)-\frac{b_0^2\left(A_0\tan\frac{t}{2}+1\right)^2}{4A_0\left(\tan\frac{t}{2}\right)^2}dr^2.
\label{KS5}
\end{equation}
\end{widetext}
In this case also we can also consider a periodic evolution of the universe, which crosses the singularities. It begins at the singularity at $t=0$ when the scale factor $a$ is equal to zero and the scale factor $b$ is infinite, then $b$ begins decreasing and arrives to a value equal to zero at the singularity at $t=t_2=2\pi-2{\rm arctan}\frac{1}{A_0}$. Meanwhile the scale factor $a$ increases, arriving to a maximum value at $t=\pi-{\rm arctan}\frac{1}{A_0}$, then it decreases and vanish at $t=t_2$. After that the scale factor
$a$ grows infinitely until arriving to the singularity at $t=2\pi$ while the scale factor $b$ reaches its maximal value at $t=2\pi - {\rm arctan}\frac{1}{A_0}$ and vanishes at $t = 2\pi$. Then the evolution repeats itself. We can add here that in the vicinity of the singularity at $t=0$,the Kasner indices are given by Eq. (\ref{Kasner1}) while in the vicinity of the singularity at $t=t_2$, the Kasner indices are given by Eq. (\ref{Kasner5}).

\section{Concluding remarks}

It is well known that on combining the conformal transformation of the metric with the reparametrization of the scalar field, one can rewrite the action of a model with a
non-minimally coupled scalar field in a form where it becomes minimally coupled.
Such a procedure is called the transformation from the Jordan frame to the Einstein frame.
For the first time this transformation was used in paper~\cite{Wagoner}.

Many papers discuss this topic, which sometimes is described as a study of the equivalence between  frames \cite{debate,KPTVV-sing}. In a way, one can say that mathematically the procedure of the transition between the frames is well defined and can be used in  different contexts. We wish to emphasize that the physical cosmological evolutions are those seen by an observer using the cosmic (synchronous) time, which is different in different frames. Thus, evolutions in the Einstein and Jordan frames, connected by a conformal transformation and by the reparametrization of the scalar field can be qualitatively different.
In the present paper we have shown that the the static spherically or hyperbolically symmetric solutions of the Einstein equations
and their Kantowski-Sachs counterparts in the presence of the conformally coupled scalar field possess some special
regimes, which are absent for the case of a minimally coupled scalar field \cite{massless-we}. Moreover, one can see that for the models considered here there exist situations when a transition from the Einstein frame to the Jordan frame or viceversa can remove or create a singularity. Similar effects were studied  in detail for Friedmann models in paper \cite{KPTVV-sing}.
It was shown that when the universe  encounter the singularity in the Einstein frame, it is absent in the Jordan frame, because
this singularity is reabsorbed by the conformal transformation factor. Such effect is however absent in the Bianchi-I
 models and the singularities arise simultaneously in both frames \cite{Starobinsky1981,we-Bianchi,we-Bianchi1}. Let us add that the fact that the conformal transformations 
 can essentially change the geometry of the spacetime due to an effective creation of some additional matter was discussed in the paper \cite{Dabrowski}.  

In recent years there has been an intensive discussion on the possibility of the crossing of the Big Bang --- Big Crunch type singularities in cosmology \cite{debate-sing}. The main point here is that one can describe the singularity crossing if in spite of the presence
of some divergent invariants at the singularity, it is possible to establish some well-defined prescription for matching some non-singular quantities before and after the singularity. In paper \cite{KPTVV-sing} such a procedure was based on the Jordan-Einstein
frame transitions. In papers \cite{we-Bianchi,we-Bianchi1} other field reparametrizations were used. In the present paper we
have used the fact that for some special choices of the parameters, the expressions for the metric and the field are well-defined
(contain only integer degrees of some simple functions) and hence, the matching between  regions separated by a singularity arises naturally. We think that the question concerning possible generality of such a procedure deserves further investigations. On the other hand the finding some exact solutions of Einstein equations which have more complicated structure than  solutions such as Friedmann-Lema\^itre universes or Schwarzschild black holes, can be useful for both cosmology and black hole physics. 
In particular, it concerns the questions connected with the general relativistic singularities. 
Let us note that spherically symmetric solutions for  models with a minimally coupled scalar field and nonzero potential have been studied in~\cite{Bronnikov:2006qj,Bronnikov:2018vbs,Ivashchuk:2003dw,Arefeva:2018jyu}. We plan to study similar solutions in the models with non-minimal coupling in the further investigations.

\section*{Acknowledgments}
AYK, SYV, EOP and AAS are supported in part by the RFBR grant 18-52-45016.


\begin{thebibliography}{99}
\bibitem{Schwarz}
K. Schwarzschild, Sitzungsber. Pruss. Akad. Wiss. Berlin (Math. Phys.) 
{\bf 1916}, 189 (1916); {\bf 1916}, 424 (1916). 
\bibitem{Tolman}
R. Tolman, Phys. Rev. {\bf 55}, 364 (1939).
\bibitem{O-V}
J. R. Oppenheimer and G. M. Volkoff, Phys. Rev. {\bf 55}, 374 (1939).
\bibitem{massless1}
I. Z. Fisher, ZhETF {\bf 18}, 636 (1948) [gr-qc/9911008].
\bibitem{massless2}
O.Bergman and R. Leipnik, Phys. Rev. {\bf 107}, 1157 (1957).
\bibitem{massless3}
H. A. Buchdal, Phys. Rev. {bf 111}, 1417 (1959).
\bibitem{massless4}
A. I. Janis, E. T. Newman and J. Winicour, Phys. Rev. Lett. {\bf 20}, 878 (1968).
\bibitem{massless5}
A. I. Janis, D. C. Robinson and J. Winicour, Phys. Rev. {\bf 186}, 1729 (1969).
\bibitem{massless6}
M. Wyman, Phys. Rev. D {\bf 24}, 839 (1981).
\bibitem{massless7}
A. G. Agnese and M. La Camera, Phys. Rev. D {\bf 31}, 1280 (1985).
\bibitem{massless8}
B. C. Xanthopoulos and T. Zannias, Phys. Rev. D {\bf 40}, 2564 (1989).
\bibitem{Bron2}
K.~A.~Bronnikov,
  J.\ Math.\ Phys.\  {\bf 43}, 6096 (2002).
\bibitem{massless-we}
M. Gaudin, V. Gorini, A. Kamenshchik, U. Moschella and V. Pasquier,
Int. J. Mod. Phys. D {\bf 15}, 1387 (2006).
\bibitem{Bronnikov:2006qj}
 K.~A.~Bronnikov, M.~S.~Chernakova, J.~C.~Fabris, N.~Pinto-Neto and
 M.~E.~Rodrigues,
   Int.\ J.\ Mod.\ Phys.\ D {\bf 17}, 25 (2008)
   \bibitem{Bronnikov:2016wqp}
  K.~A.~Bronnikov, J.~C.~Fabris and D.~C.~Rodrigues,
  Int.\ J.\ Mod.\ Phys.\ D {\bf 25}, no. 09, 1641005 (2016)
\bibitem{Galtsov}
D..~Gal'tsov and S.~Zhidkova,
  arXiv:1808.00492 [hep-th].
\bibitem{Bronnikov:2018vbs}
  K.~A.~Bronnikov,
  Particles {\bf 1}, no. 1, 5 (2018)

\bibitem{Kant-Sachs}
R. Kantowski and R. K. Sachs, J. Math. Phys. {\bf 7}, 443 (1966).
\bibitem{Harrison}
B. K. Harrison, Phys. Rev. {\bf 116}, 1285 (1959).
\bibitem{massless-we1}
L.~Rizzi, S.~L.~Cacciatori, V.~Gorini, A.~Kamenshchik and O.~F.~Piattella,
  Phys.\ Rev.\ D {\bf 82}, 027301 (2010).
\bibitem{Jordan}
P.~Jordan, Schwerkraft und Weltall, Vieweg (Braunschweig) 1955.
\bibitem{Brans}
C.~Brans and R.~H.~Dicke,
  Phys.\ Rev.\  {\bf 124}, 925 (1961).
\bibitem{Wagoner}
R.~V.~Wagoner,
  Phys.\ Rev.\ D {\bf 1}, 3209 (1970).
\bibitem{Higgs-non-min}
F.L.~Bezrukov   and M.~Shaposhnikov,
Phys. Lett. B {\bf 659}, 703 (2008);
A.O. Barvinsky, A.Yu. Kamenshchik  and A.A. Starobinsky,
J. Cosmol. Astropart. Phys. {\bf 0811},  021 (2008);
F.L. Bezrukov, A. Magnin and M. Shaposhnikov,
Phys. Lett. B {\bf 675},  88 (2009);
A. De~Simone A, M.P. Hertzberg  and F. Wilczek,
Phys. Lett.  B {\bf 678}, 1 (2009);
A.O. Barvinsky, A.Yu. Kamenshchik, C. Kiefer, A.A. Starobinsky  and C. Steinwachs,
J. Cosmol. Astropart. Phys. {\bf 0912}, 003 (2009);
F.L.~Bezrukov, A.~Magnin, M.~Shaposhnikov  and S.~Sibiryakov,
 J. High Energy Phys. {\bf 1101},  016 (2011);
A.O.~Barvinsky, A.Yu.~Kamenshchik, C. Kiefer, A.A.~Starobinsky  and C.F.~Steinwachs,
Eur. Phys. J. C {\bf 72},  2219 (2012).
\bibitem{Bekenstein}
J.D. Bekenstein, Phys. Rev. D {\bf 11}, 2072 (1975).
\bibitem{Grib}
A.A. Grib, V.M. Mostepanenko and V.M. Frolov, Theor. Math. Phys. {\bf 37}, 1065 (1978).
\bibitem{Frolov}
V.M. Frolov, A.A. Grib and V.M. Mostepanenko, Phys. Lett. A {\bf 65}, 282 (1978).
\bibitem{Melnikov1}
V. N. Melnikov and S.V. Orlov, Phys. Lett. A {\bf 70}, 263 (1979).
\bibitem{Melnikov}
V.N. Melnikov, Sov. Phys. Dokl. {\bf 24}, 471 (1980).
\bibitem{Linde1979}
A.D.~Linde, JETP Lett. {\bf 30}, 447 (1980).
\bibitem{Starobinsky1981}
A.A.~Starobinsky,
Sov. Astron. Lett. \textbf{7}, 36 (1981).
\bibitem{KPTVV2013}
A.Yu.~Kamenshchik, E.O.~Pozdeeva, A.~Tronconi, G.~Venturi and S.Yu.~Vernov,
  Classical Quantum Gravity  {\bf 31},  105003 (2014).
\bibitem{KPTVV2015}
  A.Yu.~Kamenshchik, E.O.~Pozdeeva, A.~Tronconi, G.~Venturi and S.Yu.~Vernov,
  Classical Quantum Gravity {\bf 33}, 015004 (2016).
\bibitem{KPTVV-sing}
A.Yu.~Kamenshchik, E.O.~Pozdeeva, S.Yu.~Vernov, A.~Tronconi, and G.~Venturi,
  Phys.\ Rev.\ D {\bf 94}, 063510 (2016).
\bibitem{we-Bianchi}
A.Yu.~Kamenshchik, E.O.~Pozdeeva, S.Yu.~Vernov, A.~Tronconi, and G.~Venturi,
  Phys.\ Rev.\ D {\bf 95}, 083503 (2017).
  \bibitem{we-Bianchi1}
  A.~Y.~Kamenshchik, E.~O.~Pozdeeva, A.~A.~Starobinsky, A.~Tronconi, G.~Venturi and S.~Y.~Vernov,
  Phys.\ Rev.\ D {\bf 97}, no. 2, 023536, (2018).
\bibitem{BarceloVisser}
  C.~Barcelo and M.~Visser,
  Phys.\ Lett.\ B {\bf 466}, 127 (1999); \
  C.~Barcelo and M.~Visser,
  Class.\ Quant.\ Grav.\  {\bf 17}, 3843 (2000)
\bibitem{Bron1}
K.~A.~Bronnikov,
  Acta Phys.\ Polon.\ B {\bf 32}, 3571 (2001).
\bibitem{Bron3}
K.~A.~Bronnikov and S.~V.~Grinyok,
  Grav.\ Cosmol.\  {\bf 10}, 237 (2004).
\bibitem{Bron4}
K.~A.~Bronnikov and M.~S.~Chernakova,
  Grav.\ Cosmol.\  {\bf 11}, 305 (2005).
\bibitem{Dabrowski}
M. P. Dabrowski, J. Garecki and D. B. Blaschke, Ann. Phys. (Berlin) {\bf 18}, 13 (2009). 
\bibitem{Bron-Star}
K. A.  Bronnikov and A. A. Starobinsky, JETP Lett. {\bf 85}, 1
(2007).
\bibitem{exact}
H. Stephani, D. Kramer, M. A. H. MacCallum, C. Hoenselaers and E. Herlt, Exact
Solution to EinsteinÕs Field Equations (Cambridge University Press, 2003).
\bibitem{Kasner}
E.~Kasner,
  Am.\ J.\ Math.\  {\bf 43}, 217 (1921).
\bibitem{Harvey}
A. Harvey, Gen. Rel. Grav. {\bf 22},  1433 (1990).
\bibitem{Lif-Khal}
E.~M.~Lifshitz and I.~M.~Khalatnikov,
  Adv.\ Phys.\  {\bf 12}, 185 (1963).
\bibitem{debate}
V.~Faraoni and E.~Gunzig,
  Int.\ J.\ Theor.\ Phys.\  {\bf 38}, 217 (1999);
  R.~Catena, M.~Pietroni and L.~Scarabello,
  Phys.\ Rev.\ D {\bf 76}, 084039 (2007);
  S.~Capozziello, P.~Martin-Moruno and C.~Rubano,
  Phys.\ Lett.\ B {\bf 689}, 117 (2010);
  C.F.~Steinwachs and A.Y.~Kamenshchik,
  Phys.\ Rev.\ D {\bf 84}, 024026 (2011);
  X.~Calmet and T.C.~Yang,
  Int.\ J.\ Mod.\ Phys.\ A {\bf 28}, 1350042 (2013);
    Y.N.~Obukhov and D.~Puetzfeld,
  Phys.\ Rev.\ D {\bf 90},  104041 (2014);
G.~Calcagni, C.~Kiefer and C.F.~Steinwachs,
  J. Cosmol. Astropart. Phys. {\bf 1410}, no. 10, 026 (2014);
A.Y.~Kamenshchik and C.F.~Steinwachs,
  Phys.\ Rev.\ D {\bf 91}, 084033 (2015);
L.~Jarv, P.~Kuusk, M.~Saal and O.~Vilson,
  Phys.\ Rev.\ D {\bf 91}, 024041 (2015);
G.~Domenech and M.~Sasaki,
  J. Cosmol. Astropart. Phys. {\bf 1504}, no. 04, 022 (2015).
\bibitem{debate-sing}
I.~Bars, S.H.~Chen, P.J.~Steinhardt and N.~Turok,
  Phys.\ Lett.\ B {\bf 715}, 278 (2012);
  I.~Bars, P.~Steinhardt and N.~Turok,
  Phys.\ Rev.\ D {\bf 89},  061302 (2014);
C.~Wetterich,
  Phys.\ Rev.\ D {\bf 89},  024005 (2014);
  Phys.\ Rev.\ D {\bf 90},  043520 (2014);
P.~Dominis Prester,
  Classical Quantum  Gravity  {\bf 31}, 155006 (2014);
  P.~Dominis Prester,
  Adv.\ Math.\ Phys.\  {\bf 2016}, 6095236 (2016);
J.J.M.~Carrasco, W.~Chemissany and R.~Kallosh,
  JHEP {\bf 1401}, 130 (2014);
  R.~Kallosh and A.~Linde,
  J. Cosmol. Astropart. Phys. {\bf 1401}, 020 (2014).

\bibitem{Ivashchuk:2003dw}
  V.D.~Ivashchuk, V.N.~Melnikov and A.B.~Selivanov,
  JHEP {\bf 0309}, 059 (2003)
\bibitem{Arefeva:2018jyu}
 I.Ya.~Aref'eva, A.A.~Golubtsova and G.~Policastro,
  arXiv:1803.06764 [hep-th].


\end{thebibliography}
\end{document}